\documentclass[11pt,twoside]{article}
\usepackage{cspm-asp2006}
\usepackage{epsfig,graphicx,natbib,url}  
\usepackage{lscape}
\begin{document}

\markboth{De Pontieu et al.}{Observations and Simulations of Fibrils
  and Mottles} 
\title{High Resolution Observations and Numerical Simulations of
  Chromospheric Fibrils and Mottles} 
\author{Bart De Pontieu$^{1}$, Viggo H. Hansteen$^{2}$, Luc Rouppe van der Voort$^{2}$ ,
Michiel van Noort$^{3}$ and Mats Carlsson,$^{2}$}

\affil{$^{1}$Lockheed Martin Solar and Astrophysics Lab, Palo Alto, CA, USA\\
$^{2}$Institute of Theoretical Astrophysics, University of
  Oslo, Norway \\
$^{3}$Institute for Solar Physics of the Royal 
Swedish Academy of Sciences, Sweden\\}

\begin{abstract} 
  With the recent advent of the Swedish 1-m Solar Telescope (SST),
  advanced image processing techniques, as well as numerical
  simulations that provide a more realistic view of the chromosphere,
  a comprehensive understanding of chromospheric jets such as
  spicules, mottles and fibrils is now within reach. In this paper, we
  briefly summarize results from a recent analysis of dynamic fibrils,
  short-lived jet-like features that dominate the chromosphere (as
  imaged in H$\alpha$) above and about active region plage. Using
  extremely high-resolution observations obtained at the SST, and
  advanced numerical 2D radiative MHD simulations, we show that
  fibrils are most likely formed by chromospheric shock waves that
  occur when convective flows and global oscillations leak into the
  chromosphere along the field lines of magnetic flux concentrations.
  
  In addition, we present some preliminary observations of quiet Sun
  jets or mottles. We find that the mechanism that produces fibrils in
  active regions is most likely also at work in quiet Sun regions,
  although it is modified by the weaker magnetic field and the
  presence of more mixed-polarity. A comparison with numerical
  simulations suggests that the weaker magnetic field in quiet Sun
  allows for significantly stronger (than in active regions)
  transverse motions that are superposed on the field-aligned,
  shock-driven motions. This leads to a more dynamic, and much more
  complex environment than in active region plage. In addition, our
  observations of the mixed polarity environment in quiet Sun regions
  suggest that other mechanisms, such as reconnection, may well play a
  significant role in the formation of some quiet Sun jets.
  Simultaneous high-resolution magnetograms (such as those provided by
  Hinode), as well as numerical simulations that take into account a
  whole variety of different magnetic configurations, will be
  necessary to determine the relative importance in quiet Sun of,
  respectively, the fibril-mechanism and reconnection.
\end{abstract}

\section{Introduction}

Spicules and related flows such as mottles and fibrils dominate the
highly dynamic chromosphere (Rutten, 2007, this volume), a region in which
over 90\% of the non-radiative energy going into the outer atmosphere
to drive solar activity and space weather is deposited. Spicules are
relatively thin, elongated structures, best seen at the limb in
H$\alpha$. They are but one form of the many ``spicular'' features
that dominate the chromosphere, such as quiet Sun mottles, or active
region (AR) fibrils, both observed on the disk
\citep{depontieu-Beckers1968}.  Even though it is not clear how all of
these ``spicular features'' are related, it seems that all represent
real mass motion of chromospheric plasma to coronal heights (see,
e.g., the many EUV absorbing features in images taken with the
Transition Region and Coronal Explorer, TRACE,
\cite{depontieu-Handy+etal1999}).

Spicules (at the limb), mottles (on the quiet Sun disk) and fibrils
(in active regions) have been studied for many decades, but until
recently they have remained poorly understood, because of a lack of
high quality observations: their diameters of a few hundred km and
lifetimes of a few minutes were too close to observational limits. In
addition, theoretical models were usually highly simplified (1D) and
focused mainly on spicule-like jets in isolation from the surrounding,
driving atmosphere. Several recent advances have now enabled major
breakthroughs in our understanding of at least some of these dynamic
chromospheric jets. 

On the observational side, new telescopes such as the Swedish 1 m
Solar Telescope (SST, \cite{depontieu-scharmer2003SST}) in La Palma,
as well as advanced image processing techniques, such as the
Multi-Object Multi-Frame Blind Deconvolution method (MOMFBD,
\cite{depontieu-vanNoort05MOMFBD}), have allowed diffraction-limited
($\sim$120 km) time series in chromospheric lines (e.g., H$\alpha$) at
extremely high temporal resolution (1 second). These datasets have,
for the first time, resolved the dominant temporal and spatial
evolution of active region fibrils and quiet Sun mottles
\citep{depontieu-Hansteen+etal2006,depontieu-DePontieu+etal2007}.

Parallel to these observational advances, theoretical efforts have
also made significant progress. Several papers have explored, for the
first time, the intricate relationship between the formation of
fibrils and the photospheric magneto-convective flowfield.
\citet{depontieu-Depontieu+etal2004} used a synthesis of idealized 1D
modeling and observations to propose that photospheric oscillations
can leak into the chromosphere, where they form shocks to drive
fibrils upwards.  This idea was based on the close connection between
oscillations in upper transition region moss
\citep{depontieu-DePontieu+etal1999} and the presence of
quasi-periodic jets or fibrils in the chromosphere above active region
plage \citep{depontieu-DePontieu+etal2003b}. These ideas have now been
confirmed and expanded by \citet{depontieu-Hansteen+etal2006} and
\citet{depontieu-DePontieu+etal2007} who analyzed a
diffraction-limited time series of H$\alpha$ fibrils, and directly
compared them to advanced radiative MHD simulations. They find
excellent agreement between observations and simulations.

In this paper, we will very briefly summarize the analysis of
\citet{depontieu-Hansteen+etal2006} and
\citet{depontieu-DePontieu+etal2007} in Section
\ref{depontieu-sec:fibrils}. While this analysis explains the
formation of fibrils above active region plage, it is not yet fully
clear what role this mechanism plays in quiet Sun. To explore this
issue, we show (in Section \ref{depontieu-sec:mottles}) some very
preliminary results of an analysis of very recent SST observations of
the quiet Sun chromosphere.

\section{Active Region Fibrils}
\label{depontieu-sec:fibrils}
\begin{figure}  
  \centering
  \includegraphics[width=0.8\textwidth]{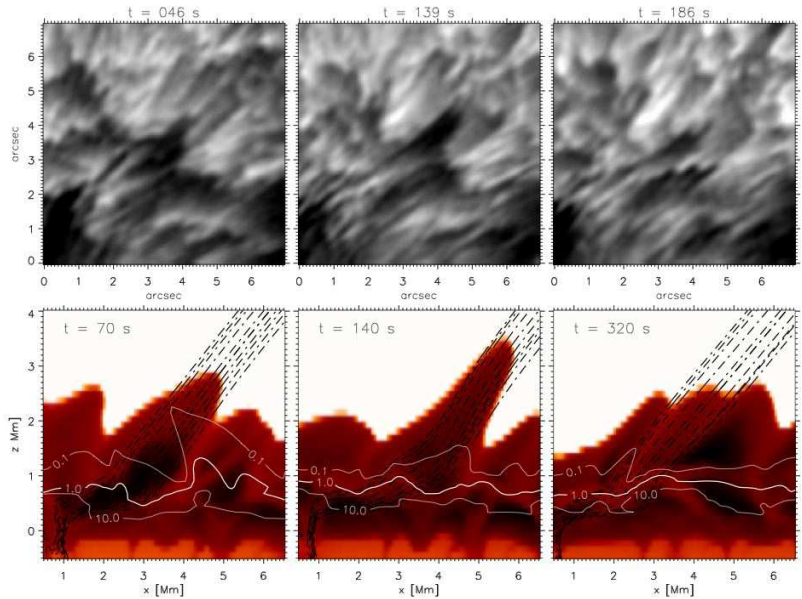}
  \caption[]{\label{depontieu-fig:fibrillife}
    Taken from \citet{depontieu-Hansteen+etal2006}. Temporal evolution
    of dynamic fibrils from H$\alpha$ linecenter observations at the
    SST (top panels), and from numerical simulations (bottom panels).
    The observations show a dark elongated feature with an upper
    chromospheric temperature of less than 10,000 K rise and fall
    within 4 minutes. Its maximum length is about 2 arcseconds.  The
    bottom panels show the logarithm of the plasma temperature T,
    which is set to saturate at log T = 4.5, from numerical
    simulations covering the upper convection zone (z$<$0) up through
    the corona (white region at the top). The vertical scale has its
    origin at the photosphere (where the optical depth
    $\tau_{500}=1$). Contours of plasma $\beta$ are drawn in white
    where $\beta=0.1, 1, 10$. The simulations show that dynamic
    fibrils ascend as a result of upwardly propagating shock waves.
    These shock waves seem to preferentially enter the corona where
    the magnetic field lines (dotted black lines) also enter the
    corona.
}
\end{figure}

\citet{depontieu-Hansteen+etal2006} and
\citet{depontieu-DePontieu+etal2007} use H$\alpha$ observations with
120 km spatial resolution and 1 s temporal resolution to study the
evolution of so-called dynamic fibrils (DFs). These fibrils are
relatively short (1,250 km on average), thin (120-700 km) and jet-like
features that occur above or in the direct vicinity of active region
plage. They are the same jets that dominate the dynamics on timescales
of minutes of the upper TR moss emission
\citep{depontieu-DePontieu+etal2003}. Extensive statistical analysis
of the H$\alpha$ timeseries reveals that most dynamic fibrils have
short lifetimes of 3 to 6 minutes, with clear differences in lifetimes
depending on the magnetic topology of the plage region. Plage regions
with more vertically oriented field are dominated by DFs with
lifetimes of 3 minutes, whereas less dense plage region or plage where
the field is more inclined from the vertical typically show DF
lifetimes of order 5 minutes. Note that what are traditionally called
H$\alpha$ fibrils typically includes a large subset of almost
horizontal loops that form a canopy-like structure that connects
neighboring plage regions or sunspots. This subset of usually much
longer fibrils is clearly more stable (in time) than the DFs. Wavelet
analysis shows that whereas the DFs are dominated by quasi-period
oscillations with periods that are identical to the DF lifetimes
(i.e., 3 and 5 mins), the horizontal fibrils show quasi-periodicity
with periods longer than 10 minutes
\citep{depontieu-Hansteen+etal2006,depontieu-DePontieu+etal2007}.

\begin{figure}
  \centering
  \includegraphics[width=0.7\textwidth]{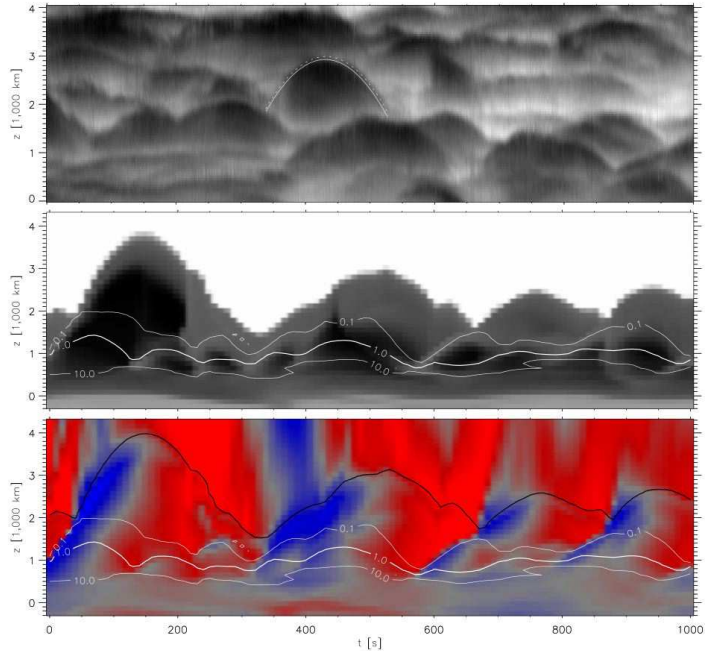}
  \caption[]{\label{depontieu-fig:parabolas}
    Taken from \citet{depontieu-Hansteen+etal2006}. Space-time plots
    of: the height of dynamic fibrils for H$\alpha$ linecenter
    observations (top panel), the logarithm of the plasma temperature
    in numerical simulations (middle panel), and simultaneous plasma
    velocity from the same simulation (bottom panel).  Most of the
    dynamic fibrils in the top panel follow a near perfect parabolic
    path in their rise and descent from the lower atmosphere.
    Parabolic paths with similar parameters are traced by fibril-like
    features in the 2D numerical simulations (middle panel). 
    Contours of plasma $\beta$ are drawn in white where $\beta=0.1, 1,
    10$.  Upward plasma velocities (blue in the
    bottom panel) show the upward propagation of the shocks that drive
    the fibril-like features along their parabolic path. This initial
    upward impulse is followed by a constant deceleration which leads
    to downward velocities (red) of roughly equal magnitude as the
    initial upward velocities. The black line in the bottom panel
    outlines the top or transition region of the simulated
    chromospheric fibrils.  }
\end{figure}

One of the major findings of \citet{depontieu-Hansteen+etal2006} and
\citet{depontieu-DePontieu+etal2007} is that almost all fibrils that
can be nicely isolated from the background or foreground of other
fibrils, follow an almost perfect parabolic path in their ascent and
descent (Figs.~\ref{depontieu-fig:fibrillife},
\ref{depontieu-fig:parabolas}). The sharp delineation of the fibril
tops is presumably related to the fact that these fibrils have sharp
transition regions at their top end. The measurements thus trace the
path of the transition region. The velocity of the TR at the beginning
of a fibril's life is between 10 and 35 km/s (projected), i.e.,
supersonic speeds. The velocity then linearly decreases with time,
reverses sign and, at the end of the fibril's life, reaches a maximum
speed that is similar in magnitude as the initial speed. The
deceleration fibrils undergo is significantly less than solar gravity,
ranging from 50 to 300 m/s$^2$. Fibrils typically have internal
structure, with various parts rising and falling with slightly
different speeds and decelerations. The widths mentioned above were
defined as the width over which the parabolic paths still look similar
in lifetime and speed.

\begin{figure}
  \centering
  \includegraphics[width=0.5\textwidth]{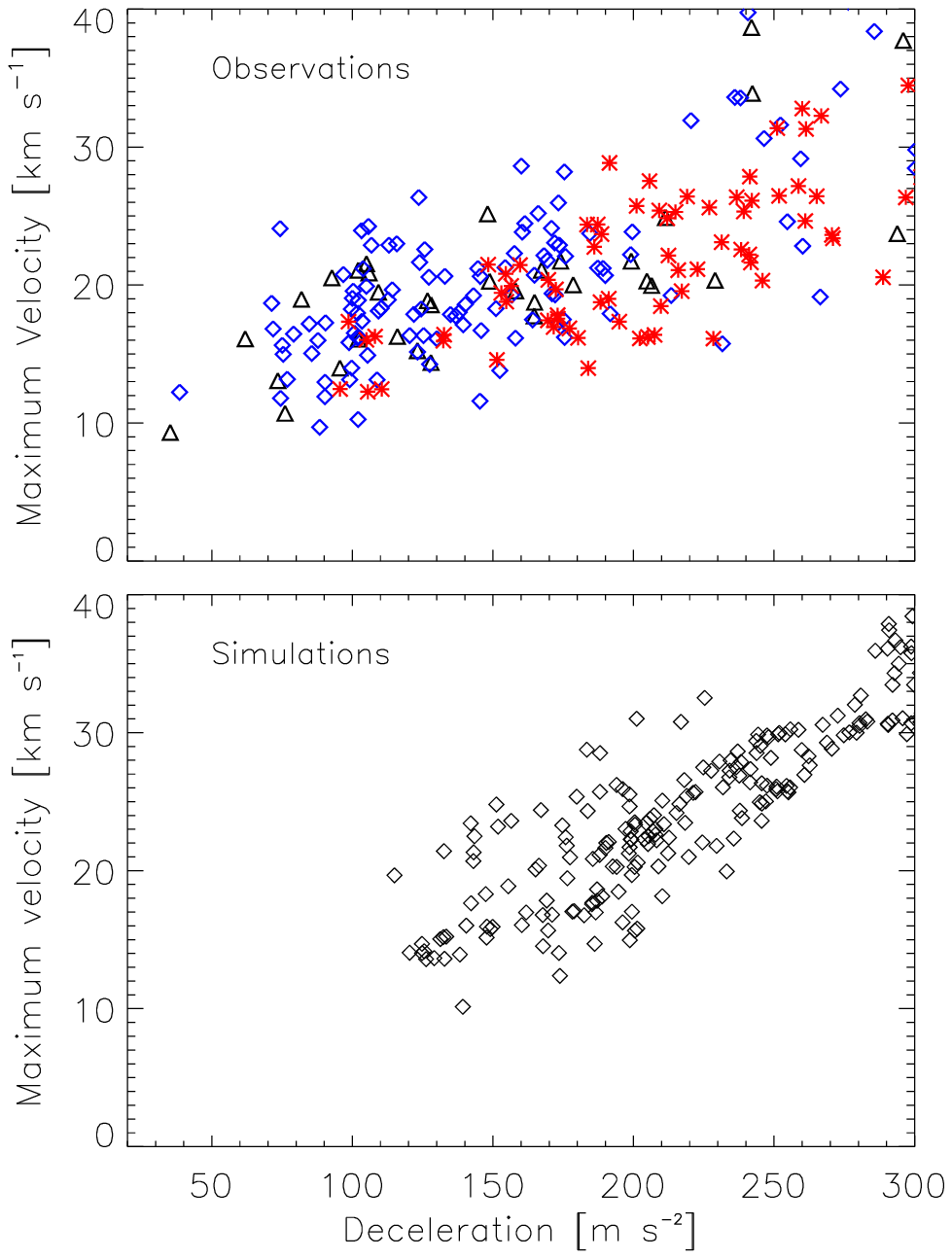}
  \caption[]{\label{depontieu-fig:correlations}
    Adapted from \citet{depontieu-Hansteen+etal2006}. Decelerations
    have been corrected by a factor of 2 (see
    \citet{depontieu-DePontieu+etal2007} for details).  Maximum
    velocities versus decelerations from 257 observed dynamic fibrils
    (top panel).  The maximum velocity and deceleration have been
    corrected for line-of-sight projection, assuming that the fibril
    is aligned with the local magnetic field as deduced from potential
    field calculations. Red stars indicate fibrils from the dense
    plage region, whereas blue diamonds indicate fibrils in the
    inclined field region. The same scatterplot (lower panel), made
    from analyzing fibril-like features in the numerical simulations,
    reveals that the simulations reproduce the observed correlation
    between these parameters, as well as reproducing the range in
    deceleration and maximum velocity.}
\end{figure}

The numerical simulations described in
\citet{depontieu-Hansteen+etal2006} and
\citet{depontieu-DePontieu+etal2007} show that fibril-like features
occur in a natural fashion in the proximity of flux concentrations, as
a result of chromospheric shocks. These shocks form when convective
flows and oscillations leak into the chromosphere along magnetic field
lines. Figure~\ref{depontieu-fig:fibrillife} shows that the jet-like
features formed in the simulations have lengths and lifetimes that are
similar to the observed fibrils. The simulations also show that the
parabolic paths observed in H$\alpha$ fibrils are a clear sign that
these jets are caused by single slow-mode magneto-acoustic shocks.
This is evident from Fig.~\ref{depontieu-fig:parabolas}, where the
shocks are clearly seen in the xt-plot of the velocity in the
simulations (bottom panel). One of the most striking results is that
the jet-like features in the simulation not only reproduce the
parabolic paths, but also the observed correlations between the
deceleration and maximum velocity
(Fig.~\ref{depontieu-fig:correlations}). 

\begin{figure}
  \centering
  \includegraphics[width=0.5\textwidth]{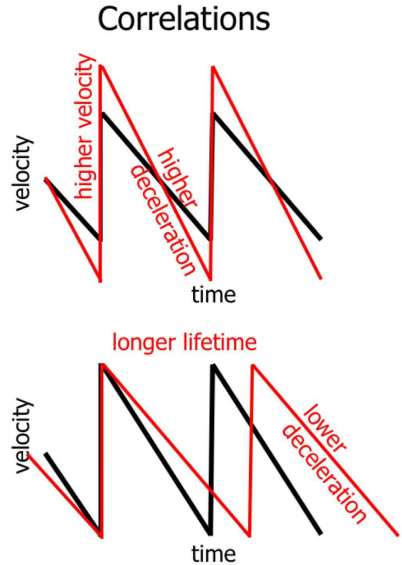}
  \caption[]{\label{depontieu-fig:correlations2}
Cartoon illustrating the cause for the correlations between the
deceleration and maximum velocity of fibrils (see
Fig.~\ref{depontieu-fig:correlations}), and between lifetime and maximum
velocity of fibrils (see \cite{depontieu-DePontieu+etal2007}). Given a
fixed lifetime (or shock wave period), higher velocity amplitudes
automatically lead to higher decelerations. For a fixed velocity
amplitude, longer wave periods lead to lower decelerations.
}
\end{figure}

This correlation can be
understood in terms of shock wave physics, as illustrated in
Fig.~\ref{depontieu-fig:correlations2}. Preliminary results of a
parameter study of detailed numerical simulations indicate that this
intuitive explanation does indeed dominate the observed correlation
between velocity and deceleration (Heggland,
De Pontieu, Hansteen, in preparation).

\citet{depontieu-Hansteen+etal2006} and
\citet{depontieu-DePontieu+etal2007} find clear differences in fibril
properties for two different plage regions. They find higher
decelerations, slightly lower velocities, shorter lifetimes and
shorter lengths for a dense plage region where the magnetic field is
inclined more vertically. In a neighboring plage region, where the
field is more inclined from the vertical they find fibrils with lower
decelerations, slightly higher velocities, longer lifetimes and longer
lengths. All of these measurements are of course projected onto the
plane of the sky, which potentially leaves the possibility that these
differences are caused by projection effects. While
\citet{depontieu-Hansteen+etal2006} and
\citet{depontieu-DePontieu+etal2007} use magnetic field extrapolations
to exclude this possibility, a more straightforward consideration
directly shows that the regional differences are not caused by
projection effects, but are rather a sign of real physical differences
between the two regions. Let us consider whether there exists a
viewing angle that would cause all differences to disappear between both
regions. This is clearly impossible, for two reasons. Firstly, the
lifetimes are very different between both regions. Lifetimes are very
well defined, and independent of viewing angle. Secondly, let us
assume a difference in viewing angle is causing the other differences.
Assume that the dense plage region is more tilted towards the observer
than the other plage region. In such a case the projected lengths,
velocities and decelerations would be lower than those measured along
the axis of the fibril. In other words, such a viewing angle
difference could resolve the difference in lengths and velocity, but
would render the difference in deceleration even worse than it already
is. This shows that projection effects cannot explain these
differences. In addition, the correlations between various parameters
that were found are also independent of projection angle, since both
parameters would be corrected by the same factor.

The fact that fibrils are caused by single shock waves is apparently
confirmed by the work of Langangen et al. (this volume), who use
H$\alpha$ spectra from the SST to determine velocities and
decelerations of dynamic fibril features. Similar shock-like features
also seem to be present in the data presented by Cauzzi et al. (this
volume), although further analysis is necessary to confirm this.

Future work will need to involve more advanced radiative transfer
calculations for H$\alpha$ in the numerical simulations. Such
calculations could provide insight into why the horizontal fibrils are
generally not as dynamic as the DFs, and whether the fact that those
fibrils are along loops that most often do not have a transition
region plays an important role in its more stable behavior.

\section{Quiet Sun Mottles}
\label{depontieu-sec:mottles}

\begin{figure}
  \centering
  \includegraphics[width=\textwidth]{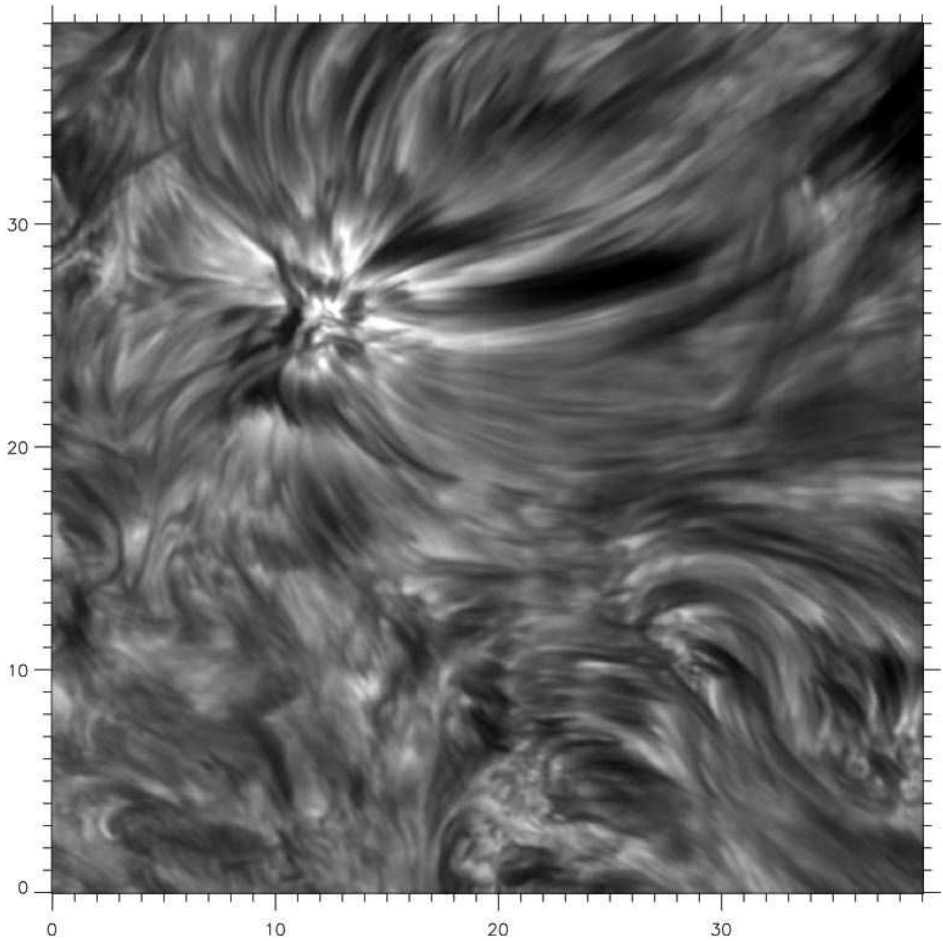}
  \caption[]{\label{depontieu-fig:qs_overview}
    Diffraction-limited image taken at the SST on 18-June-2006 showing
    H$\alpha$ linecenter of a quiet region. Note the long and curved
    dark mottles emanating from a network region around (10\arcsec,
    25\arcsec). When tracing loops from the network region to the
    upper right, a pattern below the loops becomes visible because of
    lower opacity in the loops. This is the same pattern that is
    visible in the internetwork regions that are visible in the lower left
    of the image. }
\end{figure}

Preliminary analysis of data from 18-June-2006 shows that the quiet
Sun chromosphere as imaged in a diffraction-limited, 5 s cadence
H$\alpha$ linecenter timeseries, is much more complex than similar
timeseries for active regions. The more complex topology of the quiet
Sun with its predominantly mixed polarity magnetic fields leads to a
much more diverse appearance of dark and bright features (see
Fig.~\ref{depontieu-fig:qs_overview}). Active regions are dominated by
long horizontal fibrils and short dynamic fibrils.  Quiet Sun has
equivalent features: long horizontal dark mottles and short dynamic
mottles, that both connect to network regions.  In addition, quiet Sun
shows many short, highly curved and highly dynamic features that do
not seem to be associated with network, but mostly appear in the
internetwork (see the area around 21\arcsec, 4\arcsec~in
Fig.~\ref{depontieu-fig:qs_overview}). These internetwork features can
often be seen underneath the canopy-like long horizontal mottles, so
they seem to be formed at lower heights than the ``canopy''. The
internetwork features seem to be associated with the underlying
granular dynamics.  In this section we will focus on the
network-associated features.

\begin{figure}
  \centering
  \includegraphics[width=0.7\textwidth]{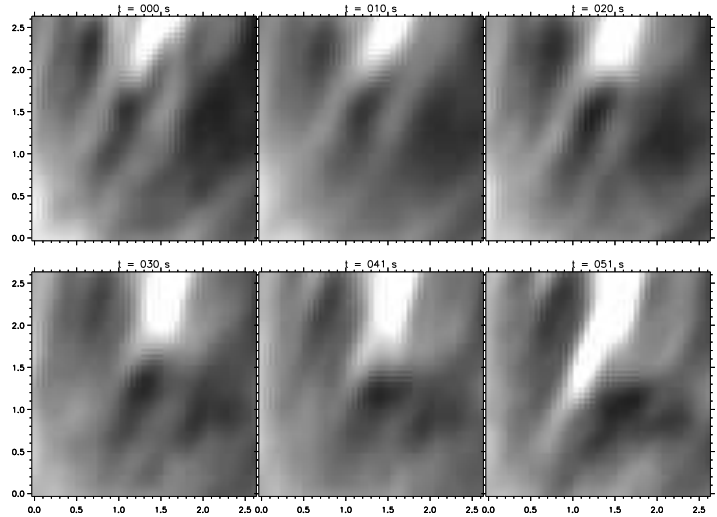}
  \caption[]{\label{depontieu-fig:mottlelife}
  Temporal evolution of a dark mottle that undergoes transverse motions from
uper left to lower right. The average velocity with which the mottle
is shifted in the direction transverse to its own axis is of order 13
km/s. Units of x and y are in arcseconds.}
\end{figure}

Contrary to active regions, it is generally extremely difficult to
track single QS mottles during their lifetime. One reason for this
difficulty is that many mottles lack the sharp ``top'' end that
dominates the appearance of dynamic fibrils in active regions. More
importantly, most mottles undergo not only up and downward motions
along the direction of the magnetic field, but also significant
motions transverse to the magnetic field, as well as fading or dimming
during their lifetime.  The transverse motions are often well
organized or coherent over several arcseconds, so that a whole batch
of almost parallel mottles seem to undergo rocking or rotating motions
during their lifetime. The apparently changing (with time) opacity as
well as the transverse motions often lead to line-of-sight
superposition, which renders unique identification throughout the
lifetime of the mottle often quite challenging.
Figure~\ref{depontieu-fig:mottlelife} shows an example of the
transverse motion a short mottle (associated with network) undergoes.
While its top end is initially at (1\arcsec, 1.7\arcsec), the whole
mottle moves to the lower right, and the top ends at (1.7\arcsec,
1.1\arcsec) after 51 seconds. The average speed transverse to the
mottle axis is of order 13 km/s. Many mottles generally undergo some
transverse motion, with typical velocities between 5 and 30 km/s,
although larger apparent velocities are also present. It is
interesting to note that turning or rotating motions have also been
observed in active region fibrils by Koza et al. (this volume). We
should also note that visual inspection of movies indicates that what
appears as a transverse motion may sometimes be an artefact of
complicated radiative transfer effects combined with coherent
wave-driven motion with phase delay between different (parallel) field
lines. For example, it is possible that some apparently transverse
motion occurs because brightening and darkening of neighboring field
lines is coherent, but with phase delays between neighboring field
lines. Such coherence and phase delays would not be surprising, since
it has also been observed in active region fibrils
\citep{depontieu-Hansteen+etal2006,depontieu-DePontieu+etal2007}.

\begin{figure}
  \centering
  \includegraphics[width=0.45\textwidth]{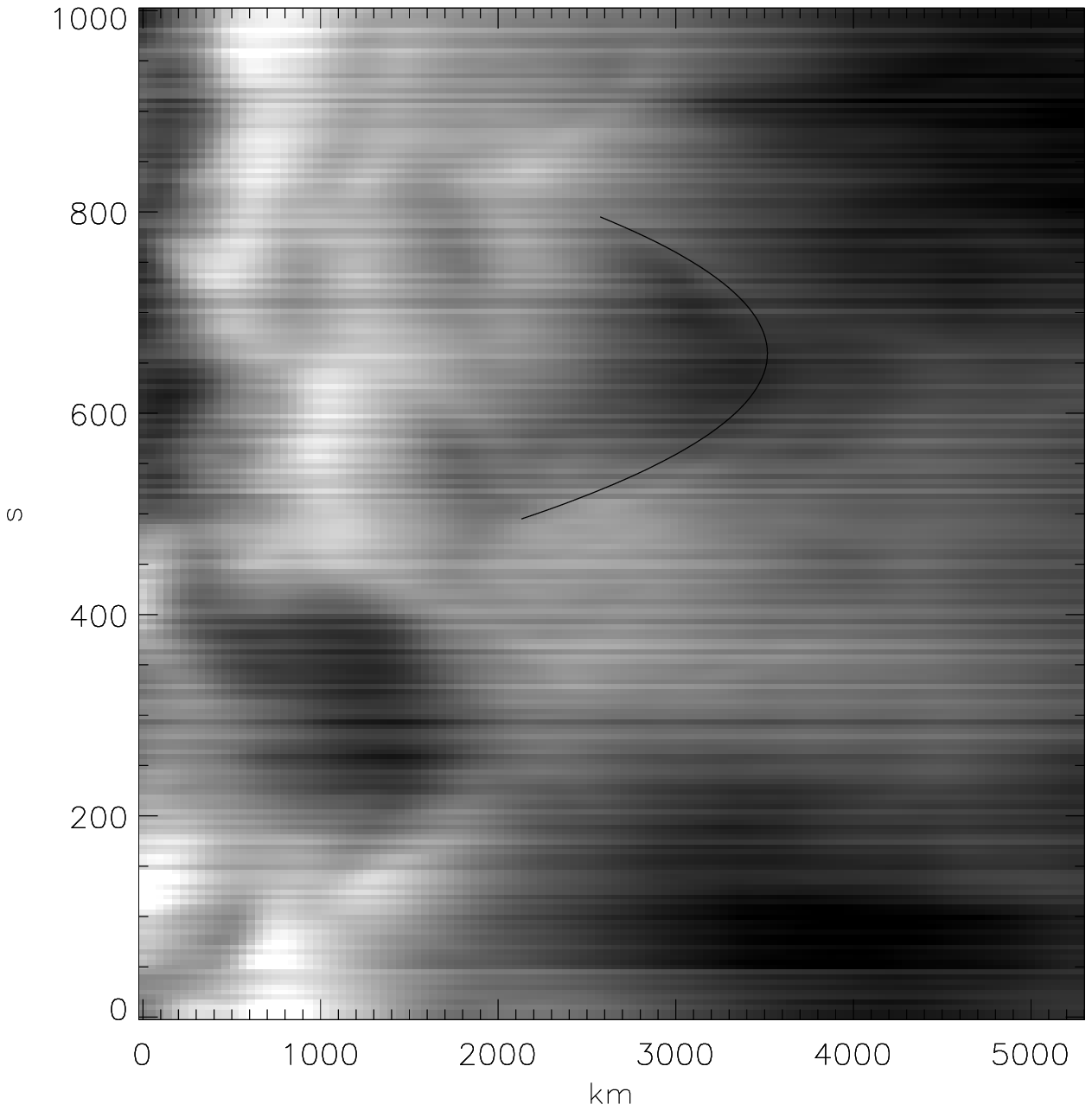}
  \hfill
  \includegraphics[width=0.45\textwidth]{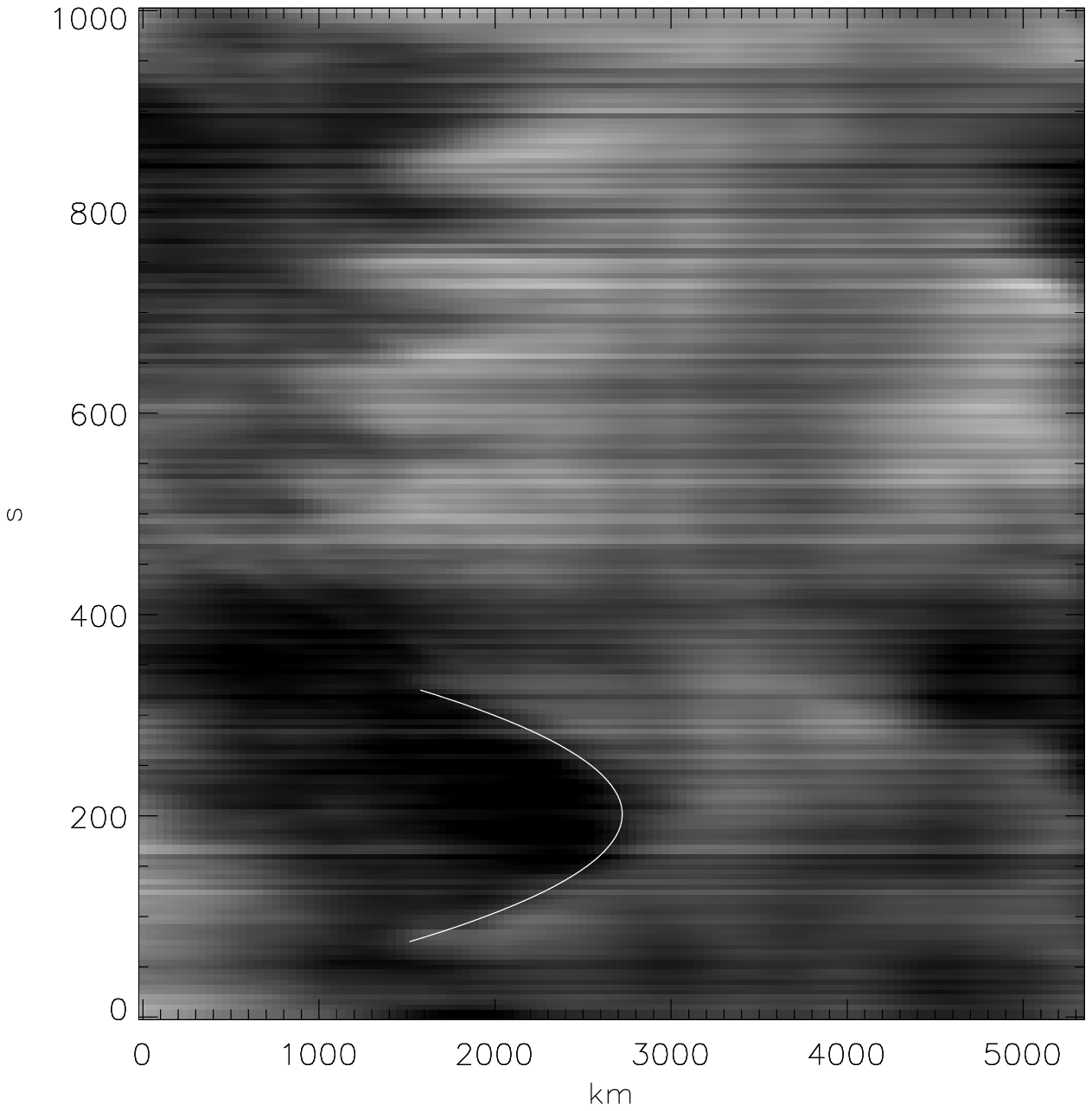}\\
  \includegraphics[width=0.45\textwidth]{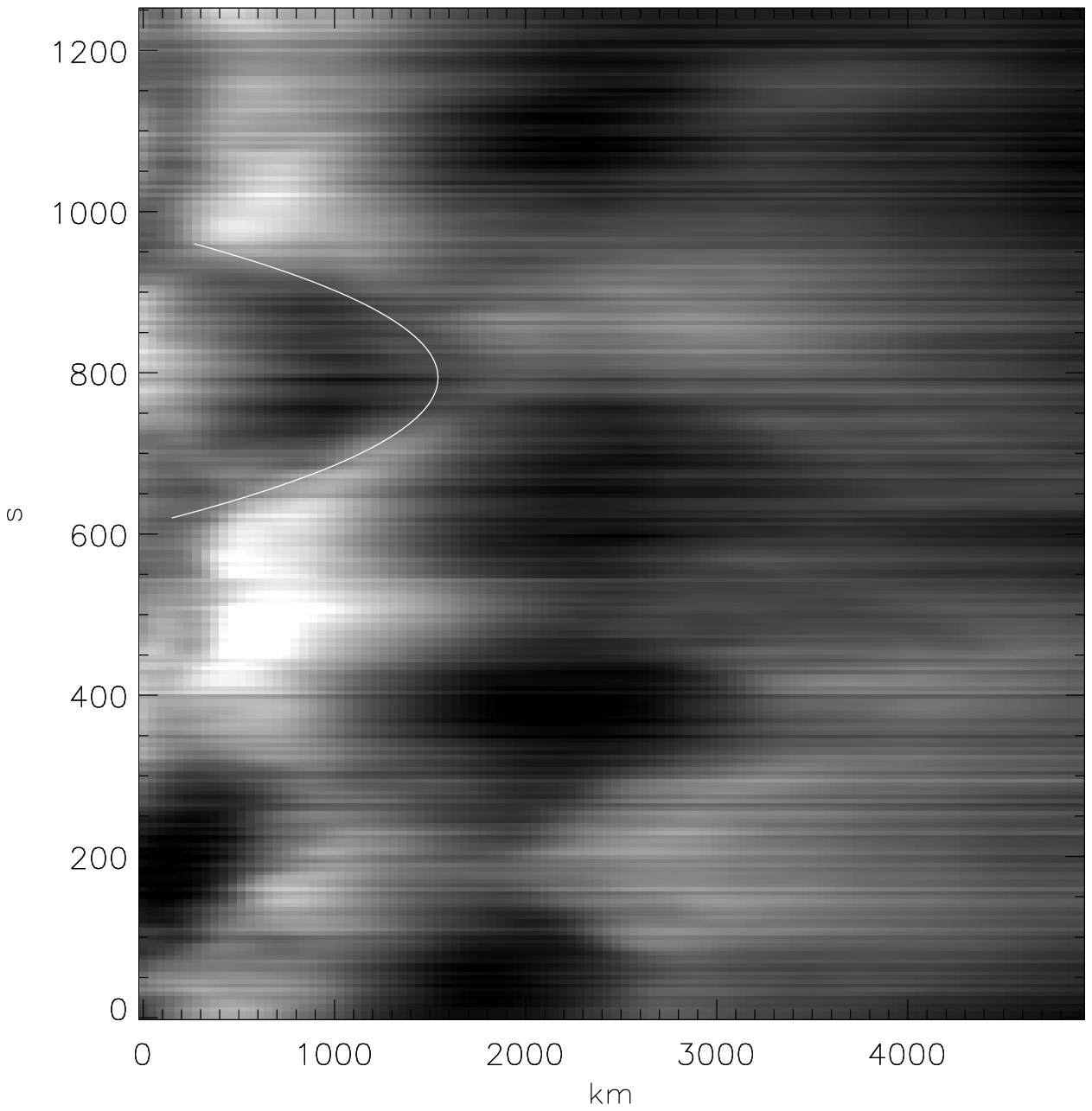}
  \hfill
  \includegraphics[width=0.45\textwidth]{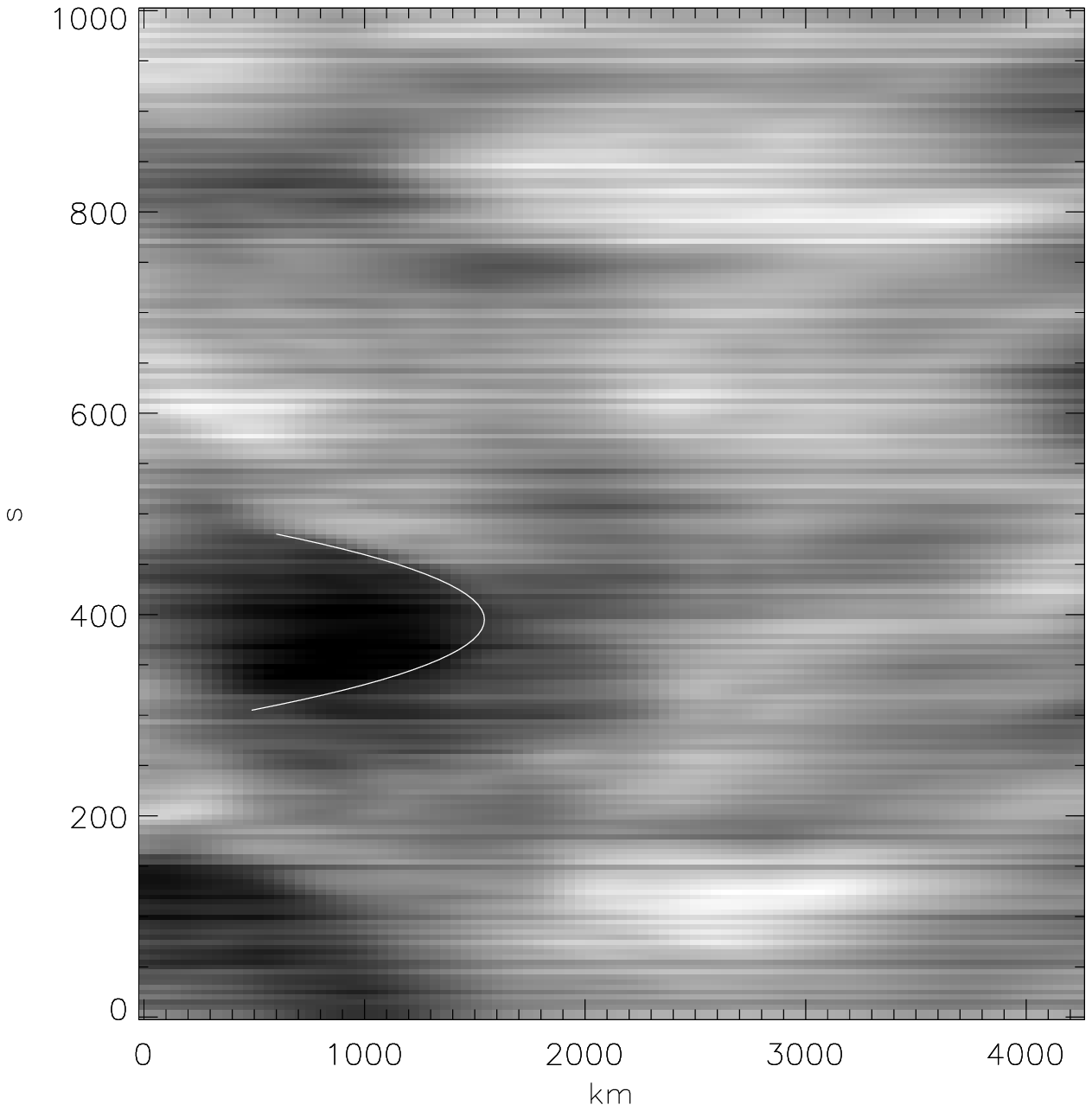}
  \caption[]{\label{depontieu-fig:qs_parabolas}
    Space-time ('xt') plots of mottles in four different areas, all
    closely associated with the network at (10\arcsec, 25\arcsec) of
    Fig.~\ref{depontieu-fig:qs_overview}. To mitigate the effects of
    transverse motion, the xt-plot has been averaged over 10 pixels
    (0.53\arcsec) in the direction perpendicular to the mottle axis.
    While most of the parabolas are not quite as clearly defined as
    those of fibrils above active region plage, it is clear that
    parabolic paths occur often for mottles that do not show much
    transverse motion, and that have a well-defined top. x-axis is in
    km, y-axis in seconds.}
\end{figure}

Despite the general difficulty with tracing individual mottles, some
mottles have a sharply-defined top end and do not show as much
transverse motion, so that xt-cuts along the mottle axis can be drawn.
In such cases, we find that the mottle top often undergoes a parabolic
path, similar to active region fibrils. Several examples are shown in
Fig.~\ref{depontieu-fig:qs_parabolas}. Why don't all mottles show such
parabolic paths? As mentioned, many mottles are not well-defined
enough, or move too much in the transverse direction to show a
parabolic path in xt-plots. The lack of a well-defined top in many QS
mottles is most probably because much of the quiet Sun chromosphere is
not in direct thermal contact (along the field line) with a hot
transition region. Such contact does occur for dynamic fibrils in
active region, so that they do have sharply defined top-ends. It is
not a coincidence that most of the well-defined parabolas of
Fig.~\ref{depontieu-fig:qs_parabolas} are found in the region directly
above or close to the strong network region, where the field is not
quite as inclined from the vertical, and coronal loops end. On heavily
inclined (i.e., almost horizontal) loops, the mottle tops are not as
well defined, presumably because there is no plasma of coronal/TR
temperatures anywhere along such loops. Preliminary comparisons with
the numerical simulations of \citet{depontieu-Hansteen+etal2006}
indicate that heavily inclined loops with apex heights less than 2 Mm
typically remain chromospheric in temperature from one end to the
other end. It is tempting to speculate that on such loops the
definition of a mottle top (or end) is no longer determined by the
steep temperature increase of the TR (and accompanying steep decrease
in opacity), but directly related to the complicated non-LTE line
formation of the H$\alpha$ line. Comparisons with detailed H$\alpha$
non-LTE radiative transfer calculations will be necssary to confirm
this suspicion.

\begin{figure}
  \centering
  \includegraphics[width=0.8\textwidth]{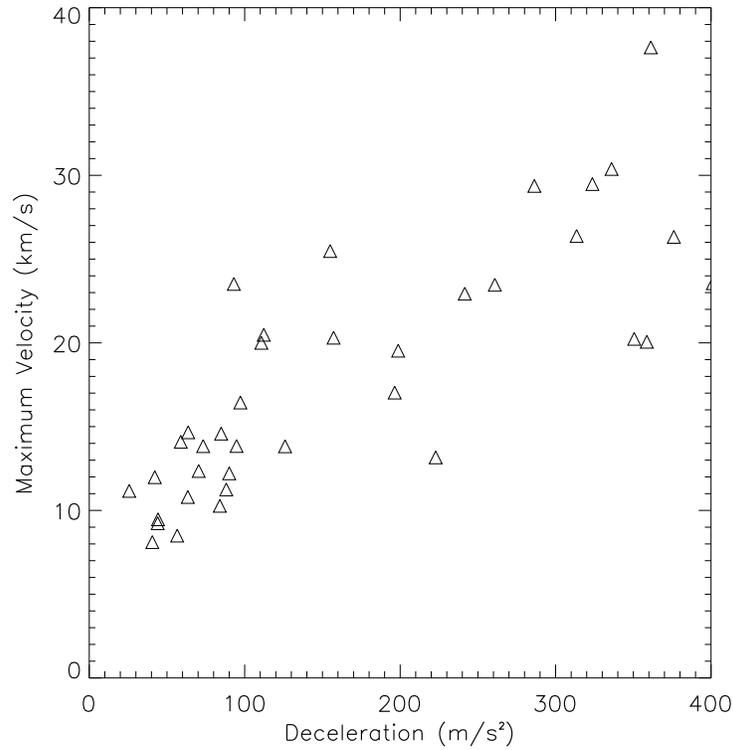}
  \caption[]{\label{depontieu-fig:qs_correlations}
    Scatterplot of maximum velocity versus deceleration for 37 quiet
    Sun mottles. Despite the relatively low statistics, it is clear
    that a linear correlation similar to the one found for active
    region fibrils (Fig.~\ref{depontieu-fig:correlations}) is present.
    A comparison with a similar correlation plot for fibril-like jets
    in numerical simulations of the chromosphere (bottom panel of
    Fig.~\ref{depontieu-fig:correlations}) strongly suggests that the
    fibril-mechanism is also important in the formation of quiet Sun
    mottles.}
\end{figure}

Regardless of the difficulties of tracking individual mottles, it is
clear that many of those that can be tracked follow parabolic paths
similar to the paths of active region fibrils. In addition, a
preliminary analysis of the correlation between deceleration and
maximum velocity of the parabolic paths of quiet Sun mottles
(Fig.~\ref{depontieu-fig:qs_correlations}) shows that the deceleration
and maximum velocity is linearly correlated.  Comparison with
Fig.~\ref{depontieu-fig:correlations} shows that the linear
correlation found for quiet Sun mottles is very similar to the one
found for active region fibrils, as well as the one found for jets in
the numerical simulations. This strongly suggests that the fibril
mechanism, i.e., leakage of convective flows and global oscillations
into the chromosphere along magnetic field lines, is also important in
the formation of jets or mottles in quiet Sun. Note that the
minimum velocities are, just like for active region fibrils, around 8
km/s, which is compatible with the chromospheric speed of sound. Such
a lower cutoff can be expected since the mottles/fibrils are driven by
chromospheric shocks.

Since these results suggest that leakage of global oscillations plays
an important role in the dynamics of the quiet Sun chromosphere, we
performed a wavelet-based oscillation analysis of two different quiet
Sun datasets. The first dataset is a diffraction-limited H$\alpha$
linecenter timeseries taken on 18-June-2006 between 11:54 and 12:42 UT
(at 5.15 s cadence). The second dataset is also diffraction-limited,
but consists of H$\alpha$ wing images ($\pm 300$ m\AA) at 18.4 s
cadence, taken on 21-June-2006 between 16:34 UT and 17:28 UT.

\begin{figure}
  \centering
  \includegraphics[width=0.84\textwidth]{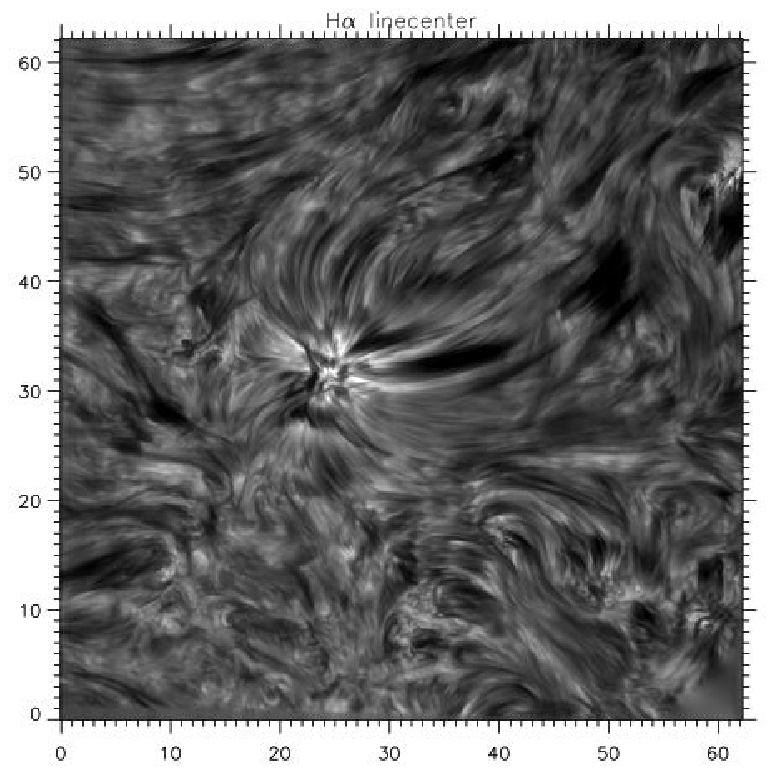}
  \hfill
  \includegraphics[width=0.84\textwidth]{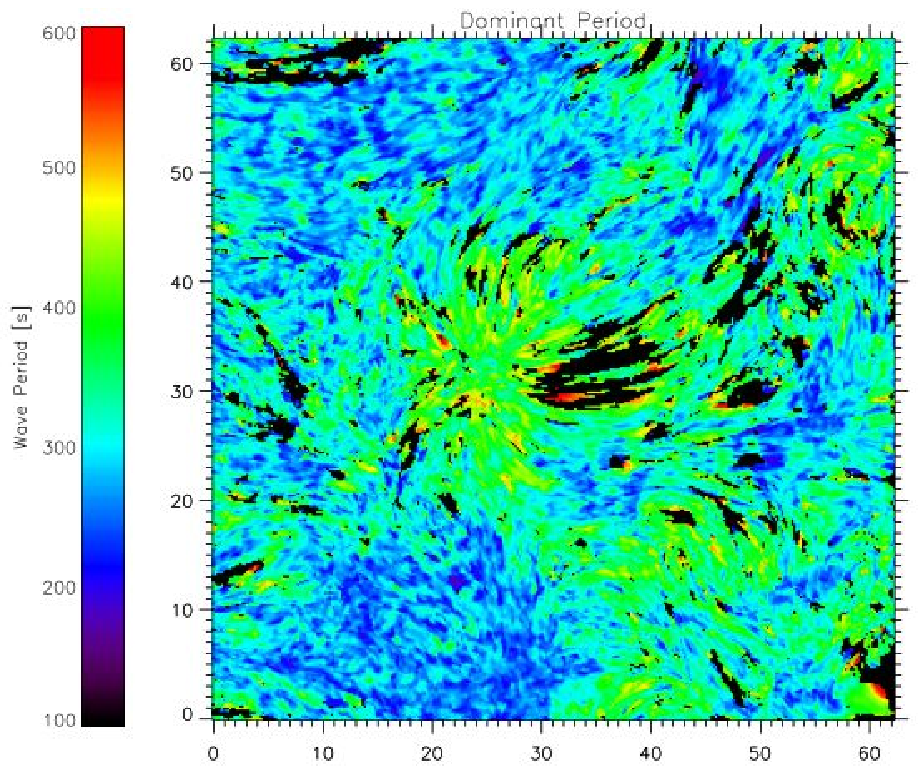}\\
  \caption[]{\label{depontieu-fig:qs_oscillations1}
    Results from a wavelet analysis of a 48 minute long H$\alpha$ 
    linecenter timeseries of quiet Sun taken on 18-June-2006.  The
    left panel shows a snapshot of the region, whereas the right panel
    illustrates for each location which wave period dominates, i.e.,
    contains the highest number of wavepackets with significant power.
  }
\end{figure}

A preliminary analysis of the dominant oscillatory behavior of the
quiet Sun chromosphere, as imaged in high-resolution H$\alpha$
linecenter timeseries, is shown in
Figure~\ref{depontieu-fig:qs_oscillations1}. This is a map of the
dominant wave period for each location of the quiet Sun chromosphere
(H$\alpha$ linecenter) that shows significant wavelet power for at
least two wave periods. The mottles and loops emanating from the
network region around (25\arcsec, 30\arcsec) are all dominated by
oscillatory behaviour with periods around 5-7 minutes, with some of
the longer and lower-lying loops dominated by periods of up to 10
minutes. This finding is compatible with the idea that leakage of
global oscillations from the photosphere (with dominant periods around
5 minutes) are important in the formation and dynamics of
network-associated mottles.  The internetwork regions, such as the
lower left region, show periods that are closer to 3 minutes. This
suggests that internetwork regions (where the field is not as
dominant) are dominated by waves with periods around the chromospheric
acoustic cutoff period of 3 minutes. It is interesting to note that
the 3 min power is also often visible in regions where the long,
low-lying loops, e.g. around (45\arcsec, 45\arcsec), that are
typically dominated by longer periods, start to become transparent.
This suggests that in such regions lower opacity of the overlying
loops allows glimpses of the internetwork dynamics and oscillations
underneath.

\begin{figure}
  \centering
  \includegraphics[width=0.71\textwidth]{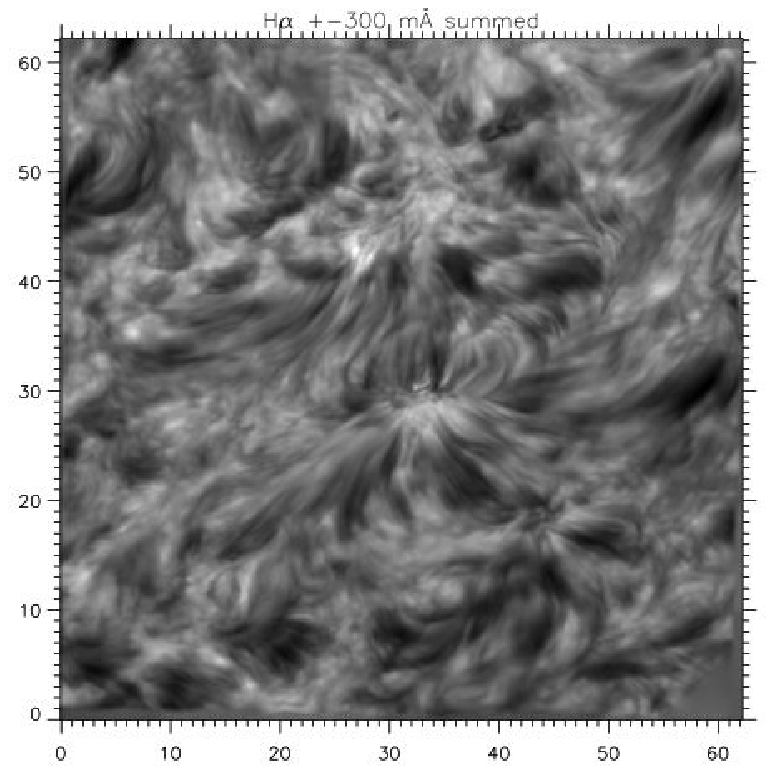}
  \hfill
  \includegraphics[width=0.71\textwidth]{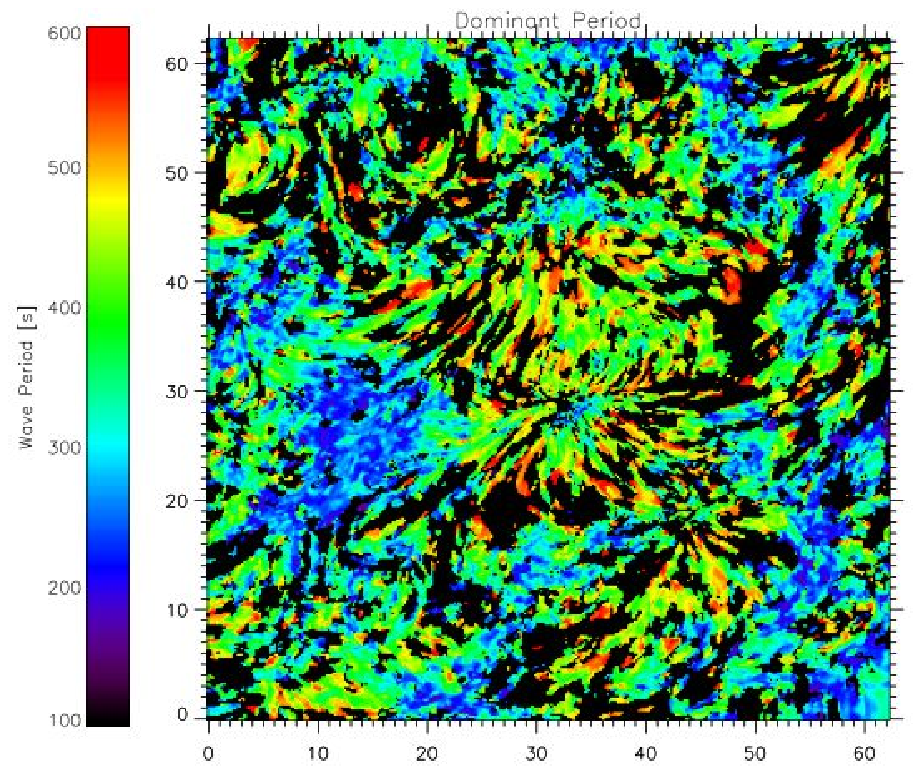}\\
  \includegraphics[width=0.71\textwidth]{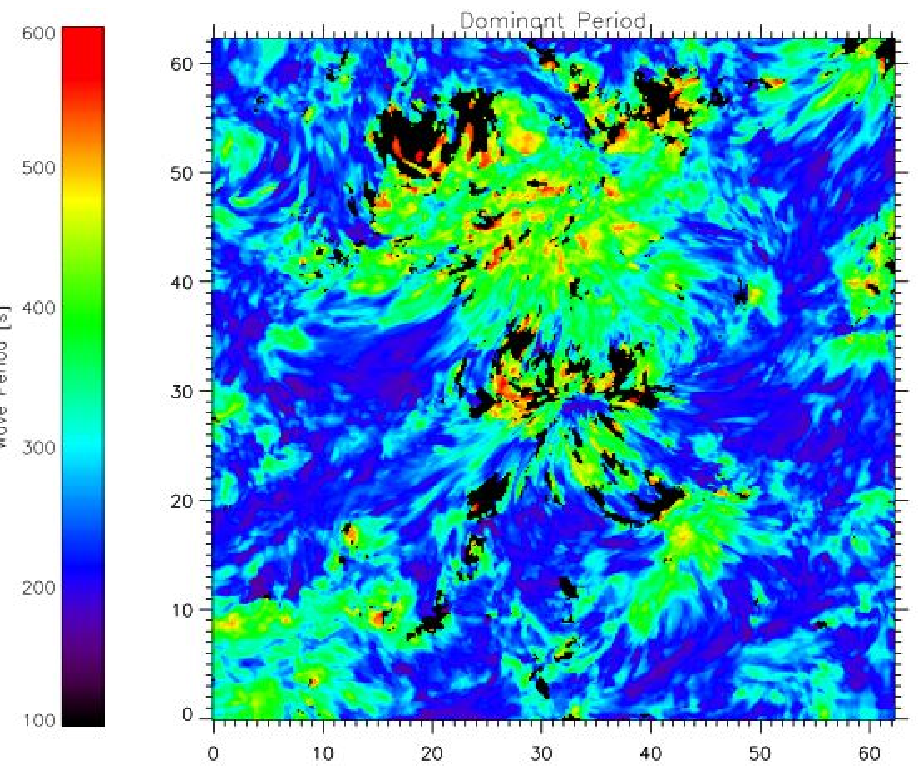}\\
  \caption[]{\label{depontieu-fig:qs_oscillations2}
    Results from a wavelet analysis of a 54 minute long H$\alpha \pm
    300$~m\AA\ timeseries of quiet Sun taken on 21-June-2006.  The top
    panel shows a snapshot of the region, the middle panel illustrates
    for each location which wave period dominates in the summed wing
    images, whereas the bottom panel shows the same for the Doppler
    signal.  }
\end{figure}

A similar picture emerges from the H$\alpha \pm 300$ m\AA\ summed
timeseries shown in the upper panels of
,Fig.~\ref{depontieu-fig:qs_oscillations2}. The network-associated
loops are dominated by periods of 5 min or longer (e.g., around
35\arcsec, 35\arcsec), whereas the internetwork regions (e.g., around
15\arcsec, 25 \arcsec) are dominated by 3 minute oscillations. This
again strongly suggests that leakage of photospheric oscillations into
the chromosphere dominates much of the dynamics of the quiet network
chromosphere.  There is a hint that right at the center of the network
regions, periods closer to 3 minutes appear to occur more often (e.g.,
at 33\arcsec, 28\arcsec). This is presumably where the field is more
vertical, so that the acoustic cutoff period reverts to its nominal 3
min value (similar to the dense plage region in
\citet{depontieu-Hansteen+etal2006}).  We also see slightly longer
periods dominating the loops connecting two opposite polarity network
regions (one centered at 35\arcsec, 30\arcsec, and one at 30\arcsec,
45\arcsec). These longer periods are reminiscent of the periods that
are seen in low-lying long fibrils that originate in sunspots or
strong plage \citep{depontieu-DePontieu+etal2007}. This may not be a
coincidence, as the quiet network regions shown here actually contain
a stronger than usual amount of magnetic flux (some tiny pores are
even visible in the continuum images, not shown).

The dominant periods in the Doppler signal are generally significantly
lower than those in the summed signal (bottom panel of
Fig.~\ref{depontieu-fig:qs_oscillations2}).  Generally, the areas
dominated by 3 min oscillations are a bit more extended in the Doppler
signal than in the summed signal, so that the ``internetwork'' regions
cover a larger fraction of the field of view. This may be partially
caused by the fact that the overlying canopy is continuously buffeted
and impinged by 3 min shocks in the internetwork (similar to the
shocks described by
\citet{depontieu-Carlsson+Stein1992,depontieu-Carlsson+Stein1994,depontieu-Carlsson+Stein1997}), an
effect which perhaps starts to dominate the velocity signal at those
locations where the canopy becomes more transparent? This could
explain why the influence of the ``network'' (i.e., 5 min signal) is
less spatially extended in the velocity signal than in the intensity
signal? Or perhaps the dominant effect is that the velocity signal
inherently contains higher frequencies than the intensity signal? The
latter is perhaps not surprising since fast-mode magnetoacoustic waves
(propagating perpendicular to the field), which predominantly appear
in the velocity signal, are ubiquitous in the numerical simulations of
\citet{depontieu-Hansteen+etal2006} and
\citet{depontieu-DePontieu+etal2007}. For example, if fast modes lift
a fibril towards the observer along the line-of-sight, such a velocity
signal would be completely absent in the intensity signal (which
defines the dark mottle). Detailed comparisons with numerical
simulations will be necessary to resolve this issue. Such simulations
will perhaps also be able to explain why the region around (30\arcsec,
50\arcsec) is dominated by 3-5 min signals in intensity, and 5-7 min
signals in velocity. The region in question is an internetwork region,
with little opacity from the overlying canopy (which may be absent
altogether at that location).  Perhaps the amplitudes in velocity are
quite low in this region, so that more long term evolution dominates
the signal compared to other internetwork regions?

In concluding, the parabolic paths, correlation between deceleration
and maximum velocity and oscillatory properties of quiet Sun mottles
strongly suggest that the mechanism that drives active region fibrils
is also responsible for the formation of at least a subset of quiet
Sun mottles. This is not surprising, since all of the ingredients of
the fibril mechanism are also present in quiet Sun: global
oscillations and convective flows that are guided into the
chromosphere along magnetic field lines. How is this mechanism
modified under quiet Sun conditions? The generally weaker magnetic
fields of the quiet Sun imply that the height of the plasma $\beta=1$
surface is generally higher than in active region plage. That implies
that the field is less rigid: flows and waves can influence the motion
of magnetic field lines up to larger heights. A less rigid field would
lead to a much more dynamic magnetic field at upper chromospheric
heights, with significantly more transverse motions. This is exactly
what we observe in our quiet Sun data. Mode coupling between different
wave modes at the plasma $\beta=1$ surface can also be expected to
play a large role in this magnetic environment
\citep{depontieu-Bogdan+etal2003}. In fact, the numerical simulations
of \citet{depontieu-Hansteen+etal2006} and
\citet{depontieu-DePontieu+etal2007} clearly show that fast-mode
magnetoacoustic waves play a significant role in the dynamics of
fibril-like jets (under weaker field conditions). These fast modes
propagate perpendicular to the fibril-axis, and can lead to a
significant transverse motion of the whole fibril-like jet. Our model
thus nicely explains the qualitative differences between fibril and
mottle dynamics.

Our observations also show many examples where significant
reorganizations of the magnetic field occur, with apparent
(un?)twisting and motions at Alfv\'enic speeds. Such reorganizations
are most probably signs of magnetic reconnection caused by the dynamic
magneto-convective driving of mixed polarity fields in the quiet Sun.
While reconnection clearly occurs often in quiet Sun, it is unclear
how much of a role reconnection plays in the formation of H$\alpha$
mottles. Taken together with the evidence presented here for a
fibril-like mechanism in quiet Sun, it seems quite possible that both
reconnection and the shockwave-driven mechanism play a role in
jet-formation in quiet Sun. It is possible that one of the reasons the
spicule problem has been so difficult to resolve is that there are
indeed multiple mechanisms at play, with perhaps a different dominant
player for jets observed in different wavelengths (e.g., visual vs.
UV). To determine which mechanism dominates where, new datasets (that
include simultaneous high resolution magnetograms) and more advanced
(3D) numerical simulations will be necessary. With the maturing of
advanced optics and post-processing techniques, the advent of larger
ground-based telescopes and satellite missions (Hinode), as well as
more advanced 3D radiative MHD simulations, a solution to these issues
is now within reach.

\acknowledgements BDP thanks the organizers of the meeting for the
opportunity to present this work. BDP was supported by NASA grants NAG5-11917,
  NNG04-GC08G and NAS5-38099 (TRACE), and thanks the ITA/Oslo group
  for excellent hospitality. VHH thanks LMSAL for excellent
  hospitality during the spring of 2006. This research was supported
  by the European Community's Human Potential Programme through the
  European Solar Magnetism Network (ESMN, contract HPRN-CT-2002-00313)
  and the Theory, Observation and Simulation of Turbulence in Space
  (TOSTISP, contract HPRN-CT-2002-00310) programs, by The Research
  Council of Norway through grant 146467/420 and through grants of
  computing time from the Programme for Supercomputing. The Swedish
  1-m Solar Telescope is operated on the island of La Palma by the
  Institute for Solar Physics of the Royal Swedish Academy of Sciences
  in the Spanish Observatorio del Roque de los Muchachos of the
  Instituto de Astrof{\'\i}sica de Canarias. 


\begin{thebibliography}{}

\bibitem[\protect\astroncite{{Beckers}}{1968}]{depontieu-Beckers1968}
{Beckers} J.~M., 1968,
  Sol. Phys.~  3, 367

\bibitem[\protect\astroncite{{Bogdan} et~al.}{2003}]{depontieu-Bogdan+etal2003}
{Bogdan} T.~J., {Carlsson} M., {Hansteen} V.~H., {McMurry} A., {Rosenthal}
  C.~S., {Johnson} M., {Petty-Powell} S., {Zita} E.~J., {Stein} R.~F.,
  {McIntosh} S.~W., {Nordlund} {\AA}., 2003,
  Ap. J.~  599, 626

\bibitem[\protect\astroncite{{Carlsson} \&
  {Stein}}{1992}]{depontieu-Carlsson+Stein1992}
{Carlsson} M., {Stein} R.~F., 1992,
  Ap. J.l~  397, L59

\bibitem[\protect\astroncite{{Carlsson} \&
  {Stein}}{1994}]{depontieu-Carlsson+Stein1994}
{Carlsson} M., {Stein} R.~F., 1994,
\newblock in M. {Carlsson} (ed.), Chromospheric Dynamics, p.~47

\bibitem[\protect\astroncite{{Carlsson} \&
  {Stein}}{1997}]{depontieu-Carlsson+Stein1997}
{Carlsson} M., {Stein} R.~F., 1997,
  Ap. J.~  481, 500

\bibitem[\protect\astroncite{{De Pontieu}
  et~al.}{1999}]{depontieu-DePontieu+etal1999}
{De Pontieu} B., {Berger} T.~E., {Schrijver} C.~J., {Title} A.~M., 1999,
  Sol. Phys.~  190, 419

\bibitem[\protect\astroncite{{De Pontieu}
  et~al.}{2003a}]{depontieu-DePontieu+etal2003b}
{De Pontieu} B., {Erd{\'e}lyi} R., {de Wijn} A.~G., 2003a,
  Ap. J.l~  595, L63

\bibitem[\protect\astroncite{{De Pontieu}
  et~al.}{2004}]{depontieu-Depontieu+etal2004}
{De Pontieu} B., {Erd{\'e}lyi} R., {James} S.~P., 2004,
  Nature~  430, 536

\bibitem[\protect\astroncite{{De Pontieu}
  et~al.}{2007}]{depontieu-DePontieu+etal2007}
{De Pontieu} B., {Hansteen} V.~H., {Rouppe van der Voort} L., {van Noort} M.,
  {Carlsson} M., 2007,
  \apj~  655, 624

\bibitem[\protect\astroncite{{De Pontieu}
  et~al.}{2003b}]{depontieu-DePontieu+etal2003}
{De Pontieu} B., {Tarbell} T., {Erd{\'e}lyi} R., 2003b,
  Ap. J.~  590, 502

\bibitem[\protect\astroncite{{Handy} et~al.}{1999}]{depontieu-Handy+etal1999}
{Handy} B.~N., {Acton} L.~W., {Kankelborg} C.~C., {Wolfson} C.~J., {Akin}
  D.~J., {Bruner} M.~E., {Caravalho} R., {Catura} R.~C., {Chevalier} R.,
  {Duncan} D.~W., {Edwards} C.~G., {Feinstein} C.~N., {Freeland} S.~L.,
  {Friedlaender} F.~M., {Hoffmann} C.~H., {Hurlburt} N.~E., {Jurcevich} B.~K.,
  {Katz} N.~L., {Kelly} G.~A., {Lemen} J.~R., {Levay} M., {Lindgren} R.~W.,
  {Mathur} D.~P., {Meyer} S.~B., {Morrison} S.~J., {Morrison} M.~D.,
  {Nightingale} R.~W., {Pope} T.~P., {Rehse} R.~A., {Schrijver} C.~J., {Shine}
  R.~A., {Shing} L., {Strong} K.~T., {Tarbell} T.~D., {Title} A.~M.,
  {Torgerson} D.~D., {Golub} L., {Bookbinder} J.~A., {Caldwell} D., {Cheimets}
  P.~N., {Davis} W.~N., {Deluca} E.~E., {McMullen} R.~A., {Warren} H.~P.,
  {Amato} D., {Fisher} R., {Maldonado} H., {Parkinson} C., 1999,
  Sol. Phys.~  187, 229

\bibitem[\protect\astroncite{{Hansteen}
  et~al.}{2006}]{depontieu-Hansteen+etal2006}
{Hansteen} V.~H., {De Pontieu} B., {Rouppe van der Voort} L., {van Noort} M.,
  {Carlsson} M., 2006,
  \apjl~  647, L73

\bibitem[\protect\astroncite{{Scharmer}
  et~al.}{2003}]{depontieu-scharmer2003SST}
{Scharmer} G.~B., {Bjelksj{\"o}} K., {Korhonen} T.~K., {Lindberg} B.,
  {Petterson} B., 2003,
\newblock in Innovative Telescopes and Instrumentation for Solar Astrophysics.
  eds. S.L. Keil \& S.V. Avakyan. Proc. SPIE.,
  Vol. 4853, p.~341

\bibitem[\protect\astroncite{{van Noort}
  et~al.}{2005}]{depontieu-vanNoort05MOMFBD}
{van Noort} M., {Rouppe van der Voort} L., {L{\"o}fdahl} M.~G., 2005,
  Sol. Phys.~  228, 191

\end{thebibliography}
\end{document}